\documentclass[
 reprint,
 amsmath, amssymb,
 aps,prd
]{revtex4-2}

\usepackage{subcaption}
\usepackage{graphicx}%
\usepackage{bm}%
\usepackage{cancel} %
\usepackage[dvipsnames]{xcolor} %
\usepackage{multirow}
\usepackage{graphicx}

\usepackage{xurl}

\usepackage[colorlinks=true,
            linkcolor=MidnightBlue,
            citecolor=ForestGreen,
            urlcolor=RoyalBlue]{hyperref}

\usepackage[nameinlink,noabbrev]{cleveref} 

\begin{document}

\preprint{APS/123-QED}

\def\apj{Astrophys.\ J.}
\def\apjl{Astrophys.\ J.\ Lett.}
\def\araa{ARA\&A}
\def\mnras{Mon.\ Not.\ R.\ Astron.\ Soc.}
\def\mnrasl{Mon.\ Not.\ R.\ Astron.\ Soc.\ Lett.}
\def\aap{Astron.\ Astrophys.}
\def\lrr{Living\ Rev.\ Relativ.}
\def\JCAP{J.\ Cosmol.\ Astropart.\ Phys.}
\def\jcap{J.\ Cosmol.\ Astropart.\ Phys.}
\def\jhep{J.\ High\ Energ.\ Phys.}
\def\cqg{Class.\ Quantum\ Grav.}
\def\prd{Phys.\ Rev.\ D\;}
\def\prc{Phys.\ Rev.\ C\;}
\def\pr{Phys.\ Rep.\ C\;}
\def\epjc{Eur.\ Phys.\ J.\ C\;}

\newcommand{\cg}[1]{{\color{black}#1}}
\newcommand{\cgg}[1]{{\color{black}#1}}
\newcommand{\st}[1]{{\color{black}#1}}

\title{
\textcolor{black}{Leave-one-out comparison of ultralight dark matter and environmental models for pulsar timing array data}
}%

\author{Shreyas Tiruvaskar}
 \email{sti50@uclive.ac.nz}
\author{Chris Gordon}%
 \email{chris.gordon@canterbury.ac.nz}
\affiliation{%
 School of Physical and Chemical Sciences, University of Canterbury, Christchurch, New Zealand
}%


\begin{abstract}
Previous work showed that ultralight-dark-matter solitons can provide
dynamical friction for supermassive black-hole binaries, suppressing
low-frequency power in the pulsar-timing-array gravitational-wave
background and constraining the particle mass and effective ultralight-dark-matter fraction.
Here we extend that analysis by comparing the predictive performance of
four models: simplified and realistic ultralight-dark-matter
implementations, a phenomenological environmental-hardening model, and
a gravitational-wave-only model. We use Bayesian leave-one-out
cross-validation on the five lowest pulsar-timing-array frequency
bins.
The phenomenological model gives the largest expected log predictive
density, but its advantage over the other models is not large compared
with the estimated standard errors. The current data therefore do not
decisively prefer one model overall. The clearest pairwise result is
within the ultralight-dark-matter framework: the simplified model
outperforms the realistic implementation in all five frequency bins.
Current pulsar-timing-array data are therefore compatible with
ultralight-dark-matter-induced low-frequency suppression, but do not yet
distinguish ultralight-dark-matter significantly from more generic environmental descriptions of
supermassive-black-hole-binary evolution.
\end{abstract}

\maketitle

\section{\label{sec:intro}Introduction}

Pulsar timing arrays (PTAs) have opened a new observational window on
nanohertz gravitational waves \cite{gwb_nanograv, gwb_ppta, gwb_epta_2023, gwb_cpta_2023, gwb_meerkat}. The stochastic gravitational-wave
background (GWB) measured by PTAs is broadly consistent with the signal
expected from a cosmological population of inspiralling supermassive
black-hole binaries (SMBHBs) \cite{gwb_smbh_Begelman_1980, Agazie_2023}. At the same time, the detailed shape of
the spectrum, especially at the lowest PTA frequencies, can depend on
the astrophysical processes that drive SMBHB orbital evolution before
gravitational-wave emission becomes dominant. These processes are
therefore relevant both to the interpretation of PTA data and to the
final-parsec problem \cite{Milosavljevic:2002ht, Milosavljevic_2003}.

In Ref.~\cite{tiruvaskar2026}, we investigated whether ultralight dark
matter (ULDM) could provide an additional source of dynamical friction
for SMBHBs. In that work, ULDM halos with central solitonic cores were
used to modify the binary hardening rate, and the resulting strain
spectra were compared with the NANOGrav 15 yr GWB data. Two ULDM
descriptions were considered: a simplified model based on an analytic
soliton profile, and a more realistic model in which the soliton profile
is modified by the gravitational influence of the SMBH binary. The
analysis showed that ULDM-induced drag can suppress the low-frequency
part of the GWB spectrum and can yield viable constraints on the ULDM
particle mass and fractional contribution to the dark matter density.

However, Ref.~\cite{tiruvaskar2026} was primarily a parameter-estimation
study. It asked what regions of the ULDM parameter space are favoured
or allowed by the PTA data, given the adopted modelling assumptions. It
did not perform a formal predictive comparison between the ULDM models
and alternative descriptions of the SMBHB population. In particular, it
did not quantify whether the simplified ULDM model, the more realistic
ULDM model, a phenomenological environmental model, or a
gravitational-wave-only model is preferred in out-of-sample predictive
accuracy. This distinction is important: a model may provide plausible
parameter constraints and visually reasonable spectra without being
clearly favoured over competing models once predictive uncertainty is
accounted for.

The purpose of the present paper is to carry out this missing model
comparison. We use Bayesian leave-one-out cross-validation (LOO-CV) to
compare the predictive performance of the ULDM models introduced in
Ref.~\cite{tiruvaskar2026} with two benchmark SMBHB models: a
phenomenological environmental model and a gravitational-wave-only
model, following the modelling framework used in the NANOGrav 15 yr
SMBHB analysis \cite{Agazie_2023}. The phenomenological model provides
a flexible description of environmental effects, while the
gravitational-wave-only model represents the limiting case in which
binary evolution is driven solely by gravitational-wave emission.

Bayesian LOO-CV provides a direct way to assess predictive accuracy.
For each data point, the model is refit with that point omitted, and
the predictive density of the omitted point is evaluated under the
leave-one-out posterior. Summing these pointwise contributions gives
the expected log predictive density (ELPD), with larger values
corresponding to better predictive performance
\cite{gelman2013bayesian,Vehtari_2016}. In this work, the LOO units are
the five lowest PTA-frequency bins used in the likelihood analysis. The
small number of LOO terms means that the resulting standard errors
should be interpreted cautiously, but the method still provides a
quantitative and transparent comparison of the models' predictive
performance on the available data.

A computationally efficient approximation to exact LOO-CV is the
Pareto-smoothed importance-sampling LOO estimate (PSIS-LOO), which
reuses samples from the full posterior \cite{Vehtari_2016,gelfand1992}.
However, the Pareto-\(\hat{k}\) diagnostics reported below indicate
that this approximation can be unreliable for some of the present
frequency-bin omissions. We therefore use exact LOO-CV as our primary
calculation. For each model and for each of the five PTA frequency bins,
we refit the model with that bin removed from the likelihood and evaluate
the predictive density of the omitted bin under the corresponding
leave-one-out posterior. This avoids relying on importance-sampling
weights in cases where a held-out frequency bin is influential.

The main result is that the phenomenological model has the largest
total exact-LOO ELPD among the models considered. However, the
differences between this model and the other models are not large
compared with the estimated standard errors. Thus, the present data do
not provide strong evidence for a statistically significant predictive
preference among all four models. The clearest pairwise result is the
comparison between the two ULDM models: the simplified ULDM model has a
larger pointwise predictive contribution than the realistic ULDM model
in all five frequency bins. This provides a quantitative follow-up to
Ref.~\cite{tiruvaskar2026}, showing that the two ULDM implementations
do not merely lead to different parameter constraints, but also differ
in their predictive performance for the observed GWB spectrum.

The structure of the paper is as follows. In
Sec.~\ref{sec:models}, we summarize the astrophysical models used in
the comparison. In Sec.~III, we describe the exact and approximate
LOO-CV methods. In Sec.~\ref{sec:results}, we present the model
comparison results, including the pointwise ELPD contributions, the
PSIS-LOO diagnostics, and the pairwise comparison between the two ULDM
models. We conclude in Sec.~\ref{sec:conclusion}.

\section{\label{sec:models}Astrophysical models}

We compare four models for the stochastic gravitational-wave background
from a cosmological population of SMBHBs. The models differ only in their physical prescription used for
binary hardening. Two are ULDM models from
Ref.~\cite{tiruvaskar2026}. The other two are benchmark SMBHB models
based on the NANOGrav 15 yr analysis \cite{Agazie_2023}.

\subsection{\label{subsec:uldm_models}Ultralight dark matter models}

ULDM consists of very light bosonic particles and can form solitonic
cores at the centres of dark-matter halos. An SMBHB moving through such
a core experiences dynamical friction, which extracts orbital energy
from the binary. This changes the binary residence time in the PTA
frequency band and therefore modifies the predicted gravitational-wave
background spectrum.

We consider the two ULDM implementations developed in
Ref.~\cite{tiruvaskar2026}. The first, denoted ``ULDM simplified'', 
uses 
a simplified prescription for the
dynamical friction. It captures the leading scaling of the ULDM drag
with the particle mass, soliton density, and binary parameters, while
neglecting the distortion of the soliton by the binary.

The second, denoted ``ULDM realistic'', uses soliton profiles modified by
the gravitational influence of the SMBH binary. This pinching of the
soliton increases the central density and can change the hardening rate
relative to the simplified model. The construction of these profiles
and their use in the SMBHB strain calculation are described in
Ref.~\cite{tiruvaskar2026}, building on the ULDM soliton-binary
dynamics of Ref.~\cite{boey2025}.

Both ULDM models use the same classes of parameters: SMBHB population
parameters, black-hole--galaxy scaling-relation parameters, and ULDM
parameters specifying the particle mass and effective ULDM fraction. In
the present work these models are treated as competing predictive
models for the PTA frequency-bin data.

\subsection{\label{subsec:phenom_model}Phenomenological environmental model}

The phenomenological model provides a flexible description of
environmental hardening without specifying a particular physical
mechanism. It parametrizes the effect of the environment on the binary
evolution through a double-power-law form, following the implementation
of Ref.~\cite{Agazie_2023}. We refer to this model as ``Phenom''.

This model is useful as a generic environmental benchmark. Unlike the
ULDM models, it is not tied to a specific dark-matter scenario, but it
can represent a broad class of low-frequency spectral modifications
caused by non-gravitational-wave hardening.

\subsection{\label{subsec:gwonly_model}Gravitational-wave-only model}

The gravitational-wave-only model assumes that SMBHB evolution is driven
solely by gravitational-wave emission. It contains no additional
hardening from stars, gas, dark matter, or other environmental effects.
We 
 refer to this model as ``GW Only''.

This model provides the baseline against which the environmental models
are compared. Differences between GW Only and the other models quantify
the extent to which the PTA data favour additional hardening or
low-frequency spectral modification.

\subsection{Model priors}
Following Refs.~\cite{Agazie_2023, tiruvaskar2026}, the astrophysical model parameters corresponding to the binary population density and black hole mass -- stellar mass relations have priors with a normal distribution. The ULDM particle mass and effective ULDM fraction have log-uniform priors. In the Phenom model, parameters related to the double-power law have uniform priors.
The exact prior assumptions for the ULDM models can be found in Table~1 of \cite{tiruvaskar2026}, and for the GW Only and Phenom models, in Table~B1 (Astrophysical Priors) of \cite{Agazie_2023}.

\section{Statistical Methods}

\subsection{Exact leave-one-out cross-validation}
\label{sec:loo}

We assess predictive performance using exact Bayesian LOO-CV. In LOO-CV, one data point is omitted,
the model is refit to the remaining data, and the predictive density
for the omitted point is evaluated. Repeating this procedure for all
data points gives a pointwise measure of out-of-sample predictive
accuracy \cite{gelman2013bayesian,Vehtari_2016}.

Let \(y=(y_1,\dots,y_n)\) denote the observed data and let
\(\theta\) denote the model parameters. Given a prior
distribution \(p(\theta)\) and likelihood \(p(y\mid\theta)\),
the posterior distribution is
\begin{equation}
    p(\theta\mid y)
    =
    \frac{p(\theta)\,p(y\mid\theta)}{p(y)}
    \propto
    p(\theta)\,p(y\mid\theta),
\end{equation}
where \(p(y)\) is the marginal likelihood
\begin{equation}
    p(y)
    =
    \int p(\theta)\,p(y\mid\theta)\,d\theta.
    \label{eq:marginal_likelihood}
\end{equation}
Although the marginal likelihood in Eq.~\eqref{eq:marginal_likelihood} can be used to form
Bayes factors, or equivalently posterior odds ratios after specifying
model prior probabilities, we do not use this as our primary model
comparison statistic. The reason is that the evidence can be sensitive to
the prior volume assigned to parameters
(e.g. Sec.~7.4 of Ref.~\cite{gelman2013bayesian} and Sec.~2 of Ref.~\cite{GordonTrotta2007}). This is important in the present application because some of
the astrophysical and ultralight-dark-matter parameters are partly
prior-limited. In particular, the effective
ultralight-dark-matter fraction is only weakly constrained
\cite{tiruvaskar2026}. In such cases, an odds ratio would compare not
only the predictive performance of the physical models, but also the
specific prior volumes adopted for poorly constrained parameters.

For this reason, we focus instead on predictive accuracy, quantified by
Bayesian leave-one-out cross-validation. This comparison asks how well
each model predicts held-out frequency-bin data after being fit to the
remaining data.

For a future or held-out observation \(\tilde y\), the posterior
predictive density is \cite{gelman2013bayesian}
\begin{equation}
    p(\tilde y\mid y)
    =
    \int p(\tilde y\mid\theta)\,p(\theta\mid y)\,d\theta\,.
    \label{eq:posterior_predictive}
\end{equation}

For exact LOO-CV, let \(y_{-i}\) denote the dataset with the
\(i^{\mathrm{th}}\) data point removed. The leave-one-out
predictive density for \(y_i\) is
\begin{equation}
    p(y_i\mid y_{-i})
    =
    \int p(y_i\mid\theta)\,p(\theta\mid y_{-i})\,d\theta.
    \label{eq:loo_predictive_density}
\end{equation}
This is the posterior predictive density of the omitted data point
under the model fit to the remaining \(n-1\) observations
\cite{gelman2013bayesian,Vehtari_2016}.

The pointwise contribution to the leave-one-out expected log
predictive density is
\begin{equation}
    \mathrm{elpd}_{\mathrm{loo},i}
    =
    \log p(y_i\mid y_{-i}),
    \label{eq:elpd_loo_i}
\end{equation}
and summing over all data points gives the exact LOO expected log
predictive density,
\begin{equation}
    \mathrm{elpd}_{\mathrm{loo}}
    =
    \sum_{i=1}^{n} \mathrm{elpd}_{\mathrm{loo},i}
    =
    \sum_{i=1}^{n} \log p(y_i\mid y_{-i}).
    \label{eq:elpd_loo_exact}
\end{equation}

If
\begin{equation}
    \theta_i^{(s)} \sim p(\theta\mid y_{-i}),
    \qquad s=1,\dots,S,
\end{equation}
are \(S\) MCMC draws from the leave-one-out posterior for the
\(i^{\mathrm{th}}\) omission, then Eq.~\eqref{eq:loo_predictive_density}
is approximated by
\begin{equation}
    p(y_i\mid y_{-i})
    \approx
    \frac{1}{S}\sum_{s=1}^{S} p(y_i\mid \theta_i^{(s)}).
    \label{eq:loo_mc}
\end{equation}
Accordingly, the Monte Carlo estimator of the pointwise LOO
contribution is
\begin{equation}
    \widehat{\mathrm{elpd}}_{\mathrm{loo},i}
    \approx
    \log\!\left(
        \frac{1}{S}\sum_{s=1}^{S} p(y_i\mid \theta_i^{(s)})
    \right),
    \label{eq:elpd_loo_i_mc}
\end{equation}
and the corresponding estimator of the total LOO score is
\begin{equation}
    \widehat{\mathrm{elpd}}_{\mathrm{loo}}
    =
    \sum_{i=1}^{n}\widehat{\mathrm{elpd}}_{\mathrm{loo},i}
    \approx
    \sum_{i=1}^{n}
    \log\!\left(
        \frac{1}{S}\sum_{s=1}^{S} p(y_i\mid \theta_i^{(s)})
    \right).
    \label{eq:elpd_loo}
\end{equation}
This is the standard Monte Carlo estimator of the Bayesian exact
LOO quantity \cite{gelman2013bayesian,Vehtari_2016}.

\textcolor{black}{We follow the computational framework used in
Ref.~\cite{Agazie_2023}. In this framework, \texttt{holodeck} is used
to simulate strain spectra from SMBHB populations for different model
parameter combinations. These simulated spectra form a training library
for an emulator. Gaussian-process interpolators are then trained on this
library so that the predicted spectra needed in the likelihood can be
evaluated rapidly at arbitrary parameter values. Following
Refs.~\cite{Agazie_2023,tiruvaskar2025}, we implement the
Gaussian-process interpolation using \texttt{ceffyl} \cite{ceffyl} and
\texttt{george} \cite{george}. The MCMC chains are generated with
\texttt{PTMCMCSampler} \cite{ptmcmcsampler}; \texttt{holodeck} is
therefore used for the population-synthesis and training-library
generation, not as the MCMC sampler.}

Following
\cite{tiruvaskar2025,tiruvaskar2026,Agazie_2023}, 
we use the five lowest-frequency bins of the NANOGrav 15 yr gravitational-wave background free-spectrum data, corresponding to the HD-w/MP+DP+CURN analysis. 
\textcolor{black}{Ref.~\cite{Agazie_2023} also considered the HD-DMGP
free-spectrum dataset. We use the HD-w/MP+DP+CURN dataset because the
strain spectrum in the fifth frequency bin is better constrained for
HD-w/MP+DP+CURN than for HD-DMGP, as shown in Fig.~1 of
Ref.~\cite{Agazie_2023}. This choice also matches the data set used in
Refs.~\cite{tiruvaskar2025,tiruvaskar2026}.}

\textcolor{black}{As a check on sensitivity to the free-spectrum data
set, Ref.~\cite{tiruvaskar2025} repeated the MCMC analysis using the
HD-DMGP data and found no significant change in the inferred results.
This previous check suggests that the present conclusions are unlikely
to be driven by the choice of HD-w/MP+DP+CURN rather than HD-DMGP,
although a full repeated LOO-CV analysis with HD-DMGP is left for future
work.}

Using the lowest five frequencies for cross-validation means \(n=5\).
In other words, the five pointwise likelihood terms correspond to the five PTA frequency-bin measurements, and these define the five LOO units in our analysis.
For each LOO
run, one of these five data points is removed from the total likelihood
and the model is refit to the remaining four points. This produces draws
\(\theta_i^{(s)} \sim p(\theta\mid y_{-i})\) for that omission. Because
only five pointwise terms contribute to the LOO sum, model-comparison
uncertainty should be interpreted with appropriate caution.

For each stored draw, we also retain the log-likelihood contribution of
the omitted point,
\begin{equation}
    \log p(y_i\mid \theta_i^{(s)}).
\end{equation}
Using these stored values, the pointwise Monte Carlo estimator can be
written as
\begin{equation}
    \widehat{\mathrm{elpd}}_{\mathrm{loo},i}
    \approx
    \log\!\left[
        \frac{1}{S}
        \sum_{s=1}^{S}
        \exp\!\left(
            \log p(y_i\mid \theta_i^{(s)})
        \right)
    \right].
\end{equation}
This expression is numerically equivalent to
Eq.~\eqref{eq:elpd_loo_i_mc}.

This procedure is exact LOO-CV in the sense that the model is
refit separately for each omitted data point. It is therefore distinct
from importance-sampling approximations to LOO, such as the raw
importance sampling and PSIS-LOO approaches discussed by
\citet{Vehtari_2016}, which instead reuse draws from the full
posterior \(p(\theta\mid y)\).

For each omission we use \(S\sim 10^5\) posterior draws. The final
LOO-CV score reported in this paper is the Monte Carlo estimate in
Eq.~\eqref{eq:elpd_loo}.

\subsection{\label{subsec:approx}Approximate LOO-CV}

The exact method described in Sec.~\ref{sec:loo} requires refitting the
model \(n\) times, once for each omitted data point. In our analysis,
the five PTA frequency-bin measurements are the pointwise likelihood
terms, and hence the five LOO units, so \(n=5\). Exact LOO-CV is
therefore computationally more expensive than an approximation that
reuses draws from the full posterior.

Approximate LOO-CV can be constructed using importance sampling
\cite{gelfand1992,Vehtari_2016}. Under the same pointwise factorization
used in Sec.~\ref{sec:loo}, the leave-one-out posterior satisfies
\begin{equation}
    p(\theta\mid y_{-i})
    \propto
    \frac{p(\theta\mid y)}{p(y_i\mid\theta)}.
    \label{eq:loo_is_identity}
\end{equation}
Thus, instead of drawing from \(p(\theta\mid y_{-i})\) separately for
each omitted data point, one may reuse draws
\(\theta^{(s)} \sim p(\theta\mid y)\) from the full posterior and
reweight them to approximate the leave-one-out predictive density.

In raw importance sampling, the importance ratios are proportional to
\begin{equation}
    r_i^{(s)}
    =
    \frac{1}{p(y_i\mid\theta^{(s)})},
\end{equation}
which can become unstable when a held-out data point is influential
\cite{gelfand1992,Vehtari_2016}. Vehtari et al.~\cite{Vehtari_2016}
therefore recommend Pareto-smoothed importance sampling (PSIS), which
stabilizes the largest importance weights by smoothing their upper tail.
Using PSIS weights \(w_i^{(s)}\), the approximate leave-one-out score is
\begin{equation}
    \widehat{\mathrm{elpd}}_{\mathrm{psis\mbox{-}loo}}
    =
    \sum_{i=1}^{n}
    \log\!\left(
        \frac{\sum_{s=1}^{S} w_i^{(s)}\,p(y_i\mid\theta^{(s)})}
             {\sum_{s=1}^{S} w_i^{(s)}}
    \right).
    \label{eq:psis_loo}
\end{equation}
This approximation avoids the \(n\) separate refits required by exact
LOO-CV, and is therefore much faster in practice.

We used the Python package \texttt{ArviZ}, described in \cite{arviz},
to implement PSIS-LOO. This method is much faster than exact LOO-CV,
but it can be unreliable for some data points if the corresponding
importance weights are poorly behaved. Its accuracy can be assessed
using the Pareto-\(\hat{k}\) diagnostic returned by the PSIS procedure
\cite{Vehtari_2016}. Roughly speaking, values \(\hat{k}<0.5\) indicate
a reliable approximation, values between about \(0.5\) and \(0.7\)
require some caution, and values \(\hat{k}>0.7\) indicate that the
approximation may be unreliable and that exact refitting is preferable
\cite{Vehtari_2016}. In our case, these diagnostics are evaluated
separately for each of the five PTA frequency bins.

In the next section, we compare the results from exact LOO-CV and
PSIS-LOO.

\section{\label{sec:results}Results}
We present the results of model comparison and the calculated ELPD values.

\subsection{Model comparison using approximate LOO-CV}

\begin{figure}[ht]
  \centering
  \includegraphics[width=0.45\textwidth]{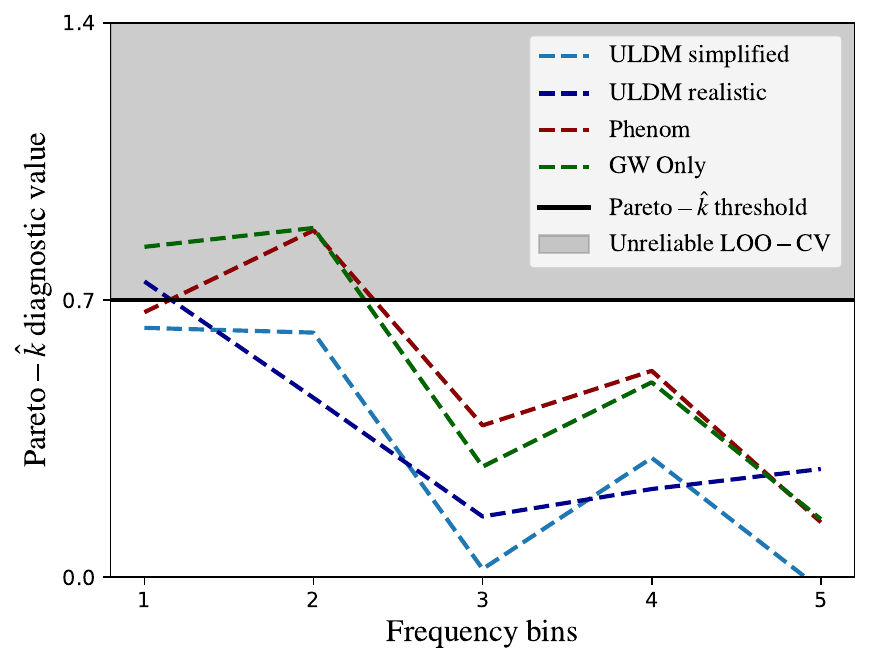}
  \caption{Pareto-\(\hat{k}\) diagnostic values for all models and all frequency bins.}
  \label{fig:pareto}
\end{figure}

For the approximate method, it is  important to assess the
Pareto-\(\hat{k}\) diagnostic, as discussed in
Sec.~\ref{subsec:approx}. The Pareto-\(\hat{k}\) values provide a
quantitative measure of the reliability of the PSIS approximation for
each pointwise term. In Figure~\ref{fig:pareto}, we present these
diagnostics for the approximate model comparison.

We can see in Figure~\ref{fig:pareto} that for the first data point,
that is, frequency bin 1, several models have Pareto-\(\hat{k}\) values
greater than \(0.7\). In contrast, for the last frequency bin, the
values for all models lie in an acceptable range. Among the models
considered here, only the ULDM simplified model has acceptable
Pareto-\(\hat{k}\) values across all five frequency bins.


Because of these high Pareto-\(\hat{k}\) values, we next use an exact LOO-CV to provide a more reliable basis for model comparison.


\subsection{Model comparison using exact LOO-CV}

In Table~\ref{tab:exact_loo}, we present the pointwise exact-LOO
contributions for each model at the five PTA frequency bins, their sum
over bins, the difference in total ELPD relative to the best-performing
model, and the corresponding estimated standard errors. Following
\citet{Vehtari_2016}, we estimate the standard error of the total LOO
score from the variation in the pointwise contributions:
\begin{equation}
    \widehat{\mathrm{SE}}\!\left(
        \widehat{\mathrm{elpd}}_{\mathrm{loo}}
    \right)
    =
    \sqrt{
        n\,V_{i=1}^{n}\!\left(
            \widehat{\mathrm{elpd}}_{\mathrm{loo},i}
        \right)
    },
    \label{eq:se_elpd_exact}
\end{equation}
where \(V_{i=1}^{n}(\cdot)\) denotes the sample variance over the
\(n\) pointwise contributions. Explicitly,
\begin{equation}
    V_{i=1}^{n}\!\left(
        \widehat{\mathrm{elpd}}_{\mathrm{loo},i}
    \right)
    =
    \frac{1}{n-1}
    \sum_{i=1}^{n}
    \left(
        \widehat{\mathrm{elpd}}_{\mathrm{loo},i}
        -
        \overline{\widehat{\mathrm{elpd}}}_{\mathrm{loo}}
    \right)^2,
\end{equation}
with
\begin{equation}
    \overline{\widehat{\mathrm{elpd}}}_{\mathrm{loo}}
    =
    \frac{1}{n}
    \sum_{i=1}^{n}
    \widehat{\mathrm{elpd}}_{\mathrm{loo},i}.
\end{equation}

For comparison with the best-performing model, we define the pointwise
difference
\begin{equation}
    \Delta \widehat{\mathrm{elpd}}_{\mathrm{loo},i}
    =
    \widehat{\mathrm{elpd}}^{\mathrm{best}}_{\mathrm{loo},i}
    -
    \widehat{\mathrm{elpd}}_{\mathrm{loo},i},
\end{equation}
so that the total difference is
\begin{equation}
    \Delta \widehat{\mathrm{elpd}}_{\mathrm{loo}}
    =
    \sum_{i=1}^{n}
    \Delta \widehat{\mathrm{elpd}}_{\mathrm{loo},i},
\end{equation}
with estimated standard error
\begin{equation}
    \Delta \widehat{\mathrm{SE}}
    =
    \widehat{\mathrm{SE}}\!\left(
        \Delta \widehat{\mathrm{elpd}}_{\mathrm{loo}}
    \right)
    =
    \sqrt{
        n\,V_{i=1}^{n}\!\left(
            \Delta \widehat{\mathrm{elpd}}_{\mathrm{loo},i}
        \right)
    }.
    \label{eq:se_delta_elpd_exact}
\end{equation}


It is also useful to quote the ELPD difference in standard-error
units.  We define the standardized LOO difference
\begin{equation}
z_{\mathrm{loo}}
\equiv
\frac{
\Delta \widehat{\mathrm{elpd}}_{\mathrm{loo}}
}{
\Delta \widehat{\mathrm{SE}}
}\, .
\end{equation}
This quantity is a descriptive measure of the separation between two
models in units of the estimated standard error.  We do not interpret it
as a formal frequentist significance or as a Gaussian $z$-score,
especially because only $n=5$ pointwise LOO terms enter the comparison.
\textcolor{black}{We estimate the uncertainty in this standardized
quantity using a delete-one-bin jackknife \cite{efron1982jackknife}.
Let \(z_{\mathrm{loo}}^{(-i)}\) denote the value of \(z_{\mathrm{loo}}\)
obtained after omitting frequency bin \(i\), and define
\(\bar z_{\mathrm{loo}}^{(\cdot)}
=n^{-1}\sum_{i=1}^{n}z_{\mathrm{loo}}^{(-i)}\). The jackknife error is
then}
\begin{equation}
{\color{black}
\sigma_{\mathrm{jack}}(z_{\mathrm{loo}})
=
\left[
\frac{n-1}{n}
\sum_{i=1}^{n}
\left(
z_{\mathrm{loo}}^{(-i)}
-
\bar z_{\mathrm{loo}}^{(\cdot)}
\right)^2
\right]^{1/2}.
}
\label{eq:zloo_jackknife}
\end{equation}





\begin{table*}[t]
\centering
\small
\setlength{\tabcolsep}{4pt}
\captionsetup{justification=raggedright,singlelinecheck=false}
\begin{tabular}{|p{2.5cm}|c|c|c|c|c|c|c|c|c|}
\hline
\multirow{2}{*}{\textbf{Model}}
& \multicolumn{5}{c|}{\textbf{ELPD per bin (\(\widehat{\mathrm{elpd}}_{\mathrm{loo},i}\))}}
& \multirow{2}{*}{\textbf{\(\widehat{\mathrm{elpd}}_{\mathrm{loo}}\)}}
& \multirow{2}{*}{\textbf{\(\Delta\widehat{\mathrm{elpd}}_{\mathrm{loo}}\)}}
& \multirow{2}{*}{\textbf{\(\Delta\widehat{\mathrm{SE}}\)}}
& \multirow{2}{*}{$z_{\mathrm{loo}}$} \\
\cline{2-6}
& \(i=1\) & \(i=2\) & \(i=3\) & \(i=4\) & \(i=5\) & & & & \\
\hline
Phenom
& -1.906 & 0.538 & -0.234 & -0.165 & -1.962
& -3.728 & 0.000  & 0.000 & -- \\

ULDM simplified
& -1.322 & 0.135 & -0.523 & -0.424 & -1.919
& -4.053 & 0.324 & 0.890 & 0.364 \\

GW Only
& -2.693 & 0.622 & -0.341 & -0.247 & -2.082
& -4.741 & 1.012 & 0.754 & 1.342 \\

ULDM realistic
& -1.556 & -0.339 & -0.918 & -0.919 & -2.256
& -5.988 & 2.259 & 1.114 & 2.028 \\
\hline
\end{tabular}
\caption{Exact LOO-CV model comparison with pointwise ELPD contributions.
\textcolor{black}{Here \(i=1,\ldots,5\) labels the five PTA frequency
bins used as LOO units, \(\widehat{\mathrm{elpd}}_{\mathrm{loo},i}\)
is the pointwise exact-LOO contribution, and
\(\widehat{\mathrm{elpd}}_{\mathrm{loo}}\) is the sum over bins.
The difference \(\Delta\widehat{\mathrm{elpd}}_{\mathrm{loo}}\) is
computed relative to the model with the largest total exact-LOO score
\(\widehat{\mathrm{elpd}}_{\mathrm{loo}}\), and
\(\Delta\widehat{\mathrm{SE}}\) is the corresponding estimated standard
error of this difference. The standardized difference is
\(z_{\mathrm{loo}}=\Delta\widehat{\mathrm{elpd}}_{\mathrm{loo}}/
\Delta\widehat{\mathrm{SE}}\). Larger ELPD values indicate better
predictive accuracy.}}
\label{tab:exact_loo}
\end{table*}

As shown in Table~\ref{tab:exact_loo}, the Phenom model has the largest
exact-LOO score and is therefore ranked highest in predictive accuracy.
However, the $z_{\mathrm{loo}}$ values are not
large. This means that the present data do not provide
strong evidence for a predictive difference between Phenom and the
other models.

\subsection{Comparison between the exact and approximate methods}
In  Table~\ref{tab:approx_loo} we show the corresponding results when using the PSIS-LOO method. Despite the larger Pareto-\(\hat{k}\) values, we can see there is only a negligible difference with the exact LOO-CV method, which was presented in Table~\ref{tab:exact_loo}. This similarity is also illustrated in Figure~\ref{fig:elpd}.

The PSIS-LOO method would have been preferable because it is substantially faster than the exact LOO-CV method, as shown in Table~\ref{tab:time}. But due to the large Pareto-\(\hat{k}\) values, we needed to check the exact LOO-CV method. Fortunately, given the scale of the problem, the exact LOO-CV times were still manageable. 







\begin{table*}[t]
\centering
\small
\setlength{\tabcolsep}{4pt}
\captionsetup{justification=raggedright,singlelinecheck=false}
\begin{tabular}{|p{2.5cm}|c|c|c|c|c|c|c|c|c|}
\hline
\multirow{2}{*}{\textbf{Model}}
& \multicolumn{5}{c|}{\textbf{PSIS-LOO per bin (\(\widehat{\mathrm{elpd}}_{\mathrm{psis\mbox{-}loo},i}\))}}
& \multirow{2}{*}{\textbf{\(\widehat{\mathrm{elpd}}_{\mathrm{psis\mbox{-}loo}}\)}}
& \multirow{2}{*}{\textbf{\(\Delta\widehat{\mathrm{elpd}}_{\mathrm{psis\mbox{-}loo}}\)}}
& \multirow{2}{*}{\textbf{\(\Delta\widehat{\mathrm{SE}}\)}}
 & \multirow{2}{*}{$z_{\mathrm{psis-loo}}$} \\
& \(i=1\) & \(i=2\) & \(i=3\) & \(i=4\) & \(i=5\) & & & & \\
\hline
Phenom
& -1.907 & 0.511 & -0.232 & -0.169 & -1.959
& -3.757 & 0.000 & 0.000 & -- \\

ULDM simplified
& -1.303 & 0.142 & -0.522 & -0.416 & -1.919
& -4.017 & 0.260 & 0.796 & 0.327 \\

GW Only
& -2.767 & 0.553 & -0.344 & -0.259 & -2.083
& -4.900 & 1.143 & 0.718 & 1.592 \\


ULDM realistic
& -1.549 & -0.335 & -0.913 & -0.922 & -2.256
& -5.975 & 2.218 & 0.989 & 2.243 \\

\hline
\end{tabular}
\caption{Approximate PSIS-LOO model comparison with pointwise predictive contributions.
\textcolor{black}{Here \(i=1,\ldots,5\) labels the five PTA frequency
bins used as LOO units,
\(\widehat{\mathrm{elpd}}_{\mathrm{psis\mbox{-}loo},i}\) is the
pointwise PSIS-LOO contribution, and
\(\widehat{\mathrm{elpd}}_{\mathrm{psis\mbox{-}loo}}\) is the sum over
bins. The difference
\(\Delta\widehat{\mathrm{elpd}}_{\mathrm{psis\mbox{-}loo}}\) is
computed relative to the model with the largest total PSIS-LOO score
\(\widehat{\mathrm{elpd}}_{\mathrm{psis\mbox{-}loo}}\), and
\(\Delta\widehat{\mathrm{SE}}\) is the corresponding estimated standard
error of this difference. The standardized difference is
\(z_{\mathrm{psis\mbox{-}loo}}=
\Delta\widehat{\mathrm{elpd}}_{\mathrm{psis\mbox{-}loo}}/
\Delta\widehat{\mathrm{SE}}\). Larger ELPD values indicate better
predictive accuracy.}}
\label{tab:approx_loo}
\end{table*}

\begin{table}[htbp]
    \centering
    \renewcommand{\arraystretch}{1.5}
    \begin{tabular}{|c|c|c|c|c|}
        \hline
        \multirow{2}{*}{\textbf{Process}} &
        \multirow{2}{*}{\textbf{Model}} &
        \textbf{\(t_{\mathrm{exact}}\)} &
        \textbf{\(t_{\mathrm{approx}}\)} &
        \textbf{Ratio} \\
        & & (min) & (min) &
        \textbf{\(\left(t_{\mathrm{exact}}/t_{\mathrm{approx}}\right)\)} \\
        \hline
        \multirow{4}{*}{MCMC}
            & ULDM simplified & 169.9 & 35.0 & 4.9 \\
            & ULDM realistic  & 160.2 & 33.4 & 4.8 \\
            & Phenom          & 84.8  & 14.4 & 5.9 \\
            & GW Only         & 21.1  & 4.1  & 5.1 \\
        \hline
        ELPD analysis & all & 0.145 & 0.071 & 2.0 \\
        \hline
    \end{tabular}
    \caption{Wall-clock times for approximately \(10^5\) samples. In
    the approximate case, one chain is used; in the exact case, five
    separate refits are required. These timing measurements were
    obtained on the University of Canterbury Research Cluster using a
    200-core CPU node with an AMD EPYC-Milan processor. We used 128
    cores while generating these chains.}
    \label{tab:time}
\end{table}

\subsection{Pairwise comparison: ULDM simplified vs.\ ULDM realistic}

\begin{figure}[ht]
  \centering
  \includegraphics[width=0.45\textwidth]{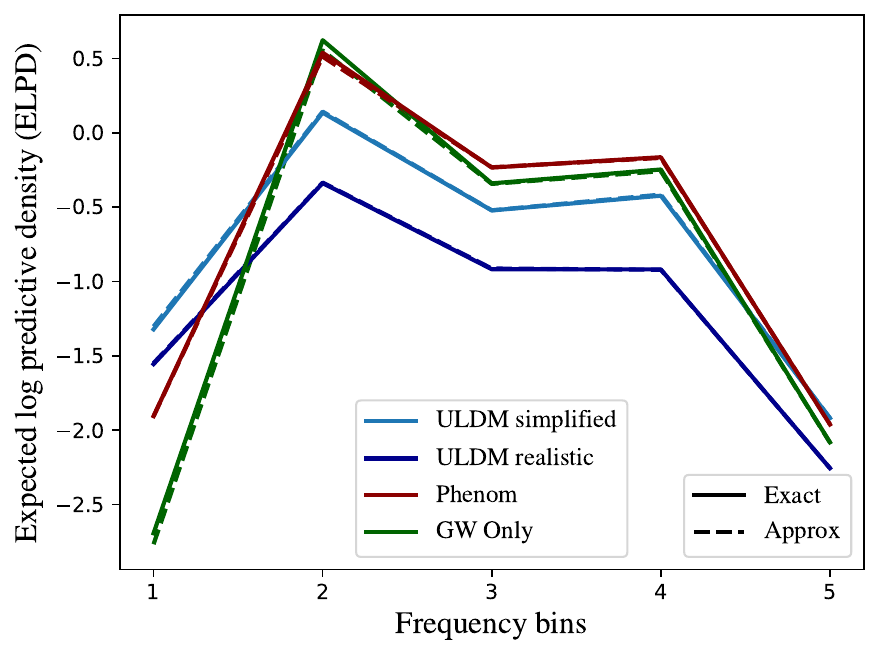}
  \caption{Pointwise predictive contributions for all models across the
  five PTA frequency bins, shown for both exact LOO-CV and approximate
  PSIS-LOO.}
  \label{fig:elpd}
\end{figure}

In Figure~\ref{fig:elpd}, we present the frequency-bin-wise predictive
contributions for all models using both exact LOO-CV and approximate
PSIS-LOO. Based on Tables~\ref{tab:exact_loo}
and~\ref{tab:approx_loo}, together with Figure~\ref{fig:elpd}, we see
that no single model is consistently preferred over all the others
across all five frequency bins. The clearest exception is the
comparison between the ULDM simplified and ULDM realistic models (blue
lines in Figure~\ref{fig:elpd}), for which the ULDM simplified model
has a larger pointwise predictive contribution in all five bins.

We also examined all six pairwise comparisons among the four models
using exact LOO-CV. Among these, the largest separation occurs between
the ULDM simplified and ULDM realistic models. For this pair, the exact
LOO difference is
\(\Delta\widehat{\mathrm{elpd}}_{\mathrm{loo}} = 1.935\) with estimated
standard error
\(\Delta\widehat{\mathrm{SE}} = 0.238\).
\textcolor{black}{Equivalently, this gives
\(z_{\mathrm{loo}} \approx 8\), with a delete-one-bin jackknife error
\(\sigma_{\mathrm{jack}}(z_{\mathrm{loo}})\approx 4\).}
\textcolor{black}{Thus, although this value of \(z_{\mathrm{loo}}\)
appears large, the small number of LOO units (\(n=5\)) makes this
standardized difference too noisy to interpret as a statistically
significant result, as reflected by the jackknife error.}

\begin{figure*}[tb!]
\begin{subfigure}{0.49\textwidth}
  \centering
  \includegraphics[width=0.95\linewidth]{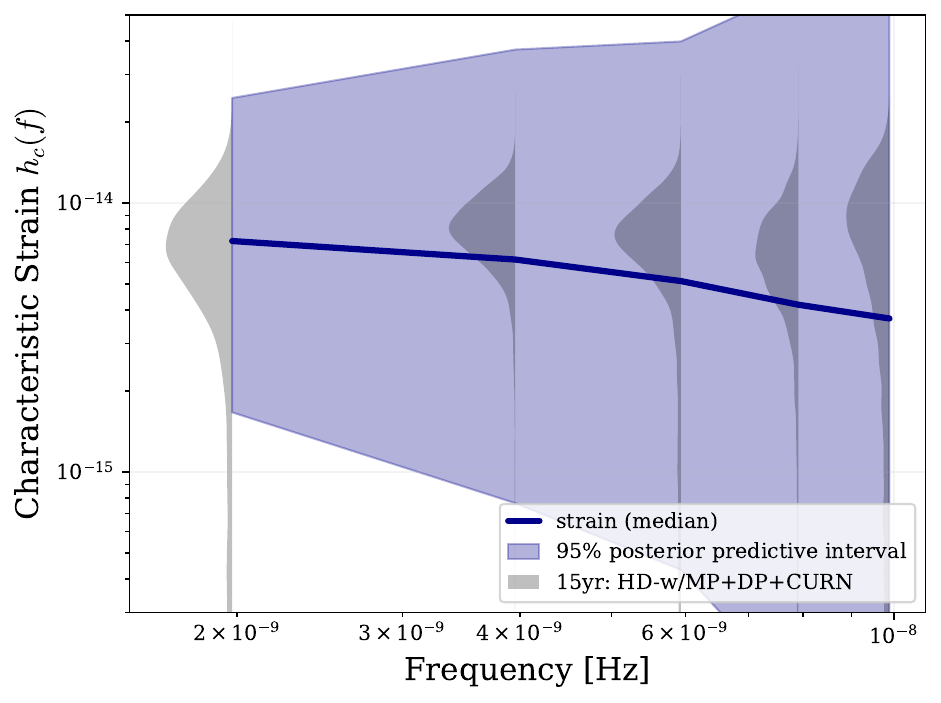}
  \label{fig:sfig2}
\end{subfigure}
\begin{subfigure}{0.49\textwidth}
  \centering
  \includegraphics[width=0.95\linewidth]{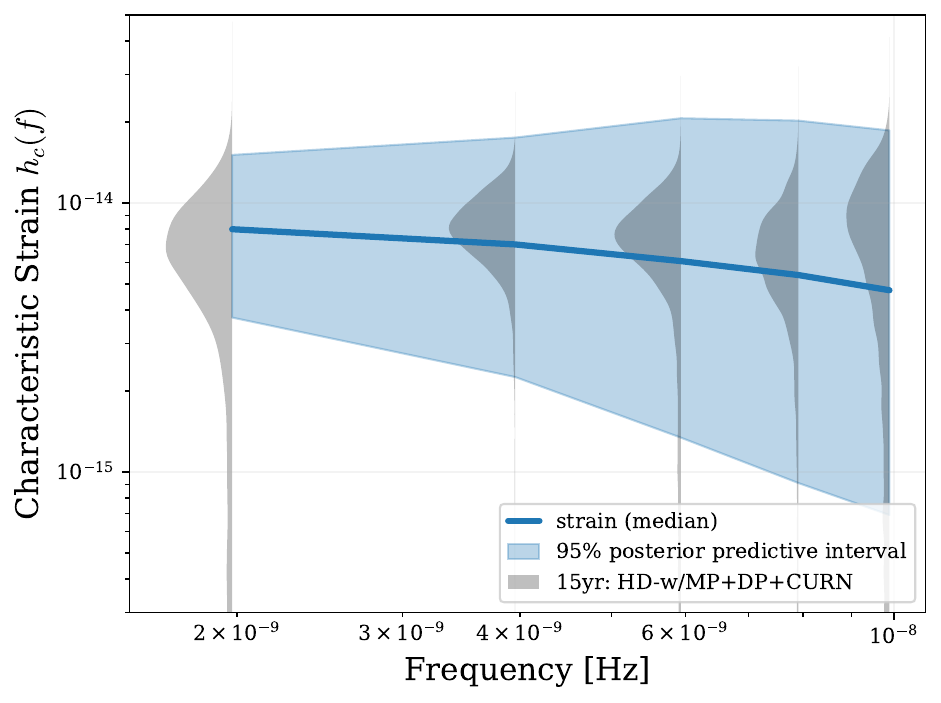}
  \label{fig:sfig1}
\end{subfigure}%
\caption{Strain spectra for all sampled MCMC parameter combinations.
We plot the median and 95\% posterior predictive interval for the strain spectra
drawn from the Gaussian-process predictive distribution. The left panel
shows the ULDM realistic model, and the right panel shows the ULDM
simplified model.}
\label{fig:strain_simple_complex}
\end{figure*}

This difference is also evident from the blue lines in
Figure~\ref{fig:elpd}. The light-blue curve for the ULDM simplified
model lies above the dark-blue curve for the ULDM realistic model in
all five frequency bins. To understand why the simplified model
performs better for the observed GWB data, we return to the MCMC chains
for both models. For each sampled parameter combination, we use the
trained Gaussian-process interpolator to obtain a predictive
distribution for the strain spectrum and then draw from that
distribution. Repeating this over \(\sim 10^5\) sampled parameter
combinations yields an ensemble of strain spectra across the five PTA
frequencies. From these, we compute the median and 95\% posterior predictive interval for each model, shown in
Figure~\ref{fig:strain_simple_complex}.

Figure~\ref{fig:strain_simple_complex} suggests that the strain
predictions for the ULDM simplified model (right panel) are more
concentrated near the observed data points (grey violins) than those of
the ULDM realistic model. By contrast, the ULDM realistic model (left
panel) yields a broader predictive distribution across the five PTA
frequencies. This provides a qualitative explanation for why the ULDM
simplified model achieves better predictive performance on the observed
GWB data.

\section{\label{sec:conclusion}Conclusion}

This work extends the ULDM analysis of Ref.~\cite{tiruvaskar2026} by
performing a predictive model comparison. That earlier study showed
that ULDM-induced dynamical friction can suppress low-frequency power in
the PTA gravitational-wave background and can yield viable constraints
on the ULDM particle mass and effective fraction. Here we asked whether
the resulting ULDM models are preferred, in predictive accuracy, over
alternative SMBHB models.

We compared four models: the simplified and realistic ULDM models of
Ref.~\cite{tiruvaskar2026}, a phenomenological environmental model, and
a gravitational-wave-only model. The comparison used Bayesian
leave-one-out cross-validation, with the five lowest PTA frequency bins
as the LOO units. 

The phenomenological environmental model has the largest total
exact-LOO ELPD. However, the differences between this model and the
others are not large compared with the estimated standard errors. Thus,
with the present five-bin dataset, we do not find strong evidence for a
decisive predictive preference among the four models. 

\textcolor{black}{The most consistent pointwise pattern is in the
comparison between the two ULDM models: the simplified ULDM model has a
larger exact-LOO contribution than the realistic ULDM model in all five
frequency bins. However, as discussed above, with only five LOO units
this standardized difference is too noisy to interpret as a
statistically significant preference.}
The PSIS-LOO calculation gives a very similar numerical difference, but due to the
Pareto-\(\hat{k}\) diagnostics, we needed to check the exact LOO-CV method.

These results clarify the status of the ULDM interpretation. Current
PTA data are consistent with ULDM-induced suppression of low-frequency
power, but they do not yet prefer ULDM over more generic environmental
descriptions of SMBHB evolution.

Future PTA datasets, with more frequency bins and smaller
uncertainties, should make this comparison more discriminating. On the
theory side, improved modelling of SMBHB evolution in ULDM solitons and
of the core--halo relation in mixed ULDM--CDM scenarios will be needed
before predictive preferences can be translated into robust physical
claims.

\section*{Acknowledgments}
 We thank Russell Boey and Richard Easther for helpful discussions.
We gratefully acknowledge support from the Marsden Fund Council grant
MFP-UOA2131, managed by the Royal Society Te Ap\={a}rangi with New Zealand
Government funding. We also acknowledge the University of Canterbury
Research Cluster facilities for providing computational resources
(\href{https://doi.org/10.18124/CANTERBURYNZ-UCRCH}
{DOI:10.18124/CANTERBURYNZ-UCRCH}, RRID:SCR\_027870).
    
\section*{Data availibility}
The code and the data used for producing the results of this paper are publicly available \citep{tiruvaskar_2026_zenodo}.

\clearpage %
\bibliography{apssamp}%

\end{document}